\documentclass[12pt,leqno]{article}

\usepackage{graphicx}
 \font\gotb eufm10 scaled \magstep1
\newcommand{\bb}{\bibitem}
\newcommand{\cc}{\cite}
\newcommand{\vp}{\varphi}
\newcommand{\sss}{\sigma}

\newcommand{\Om}{\Omega}

\newcommand{\lt}{\left}
\newcommand{\rt}{\right}

\newcommand{\F}{{\cal F}}

\newcommand{\K}{\hat K}
\newcommand{\I}{\hat I}

\newcommand{\A}{\hat A}
\newcommand{\B}{\hat B}

\newcommand{\p}{\hat p}

\newcommand{\AAA}{\hbox{\gotb A}}

\newcommand{\HHH}{\hbox{\gotb H}}

\newcommand {\QQQ}{\hbox {\gotb Q}_{\xi}}
\newcommand {\qqq}{\hbox {\gotb Q}_{\xi'}}
\newcommand {\vx}{\vp_{\xi}}
\newcommand {\vpx}{\vx(\A)}
\newcommand{\OO}{{\cal O}}
\newcommand{\BB}{\hbox{\gotb B}}
\newcommand{\bea}{\begin{eqnarray} \label}
\newcommand{\eeq}{\end{equation}}
\newcommand{\beq}{\begin{equation} \label}
\newcommand{\eea}{\end{eqnarray}}

\newcommand{\rr}[1]{(\ref{#1})}
\newcommand{\rem}{{\sc Remark}}
\newcommand{\post}{{\sc Postulate}}
 \author{D.A.Slavnov \thanks{E-mail slavnov@goa.bog.msu.ru}}

\title{Locality problem in quantum theory\thanks{Theoretical and Mathematical Physics, 155(2): 788-800 (2008)} }

   \date{}

\begin{document}
%\large

  \maketitle

 \begin{abstract}

We discuss the locality problem in relativistic and
nonrelativistic quantum theory. We show that there exists a
formulation of quantum theory that, on one hand, preserves the
mathematical apparatus of the standard quantum mechanics and, on
the other hand, ensures the satisfaction of the locality condition
for each individual event including the measurement procedure. As
an example, we consider the scattering from two slits.

\end{abstract}


\section{Introduction}

Since the famous debates between Einstein~\cc{epr,ein1,ein2} and
Bohr~\cc{bohr1,bohr2}, specialists have argued about locality in
quantum mechanics. Einstein claimed that the locality principle is
violated in quantum mechanics, which is therefore inconsistent or
at least incomplete. Bohr's main counterargument was that physical
reality and therefore locality in quantum theory cannot be
interpreted as Einstein did. The contradiction between Einstein
and Bohr was mainly due to their different views of the basic
purpose of physics. Einstein considered that the basic purpose is
to understand and describe the structure of the material world.
Bohr thought that the basic purpose is to formulate rules that one
specialist can use to communicate to another specialist what must
be done to repeat an experiment and obtain the same result.

In the years following the debates, Einstein made no progress on
his way to quantum theory. In contrast, Bohr's followers were
realizing his program with much success.

As a result, the scientific community was overwhelmingly convinced
that Bohr was right in the argument. More precisely, the vast
majority of specialists in the field of quantum mechanics prefer
not to go into the depths of this argument, considering it a
purely philosophical problem not deserving attention. But because
they successfully apply rules developed by Bohr's supporters, they
are automatically included in Bohr's camp. On the other hand, most
of them consider themselves solving the problem of describing
reality. Such a dual position cannot but affect quantum physics.
The relation between relativistic quantum field theory and
nonrelativistic quantum mechanics is vague as regards the locality
problem.

The locality property occupies the central place in quantum field
theory. The so-called Einstein locality is one of the basic
postulates in the Wightman approach~\cc{wigh}. In Bogoliubov's
approach~\cc{bog}, the causality condition is the principal
constructive element. This condition is closely related to
locality and states that any excitation in one domain of the
Minkowski space does not affect physical processes in another
spacelike separated domain. Finally, the algebraic approach
attributed mainly to Haag~\cc{haag} and Araki~\cc{araki} is
completely based on algebras of local observables, i.e.,
observables that can be measured in a bounded domain of the
Minkowski space.

The locality property is not denied in nonrelativistic quantum
mechanics when considering interactions of quantum objects. But
the situation is drastically changed when discussing the problem
of interaction between quantum objects and classical measuring
devices. The main constructive tool describing such an interaction
is the projection principle~\cc{von}. In the most cases, the
projection principle had indeed proved its usefulness when
describing the influence of a measuring device on a quantum object
in the framework of the standard mathematical apparatus of quantum
mechanics. But the physical mechanism for realizing this principle
as well as its consistency with the locality condition is still
missing.

Instead, people vaguely reason that the human brain has an
experience of describing only classical objects and therefore
cannot provide a concrete image of how a quantum object interacts
with a classical device. But if the human brain can provide a more
or less explicit picture of interactions between quantum objects,
why should it fail when providing a picture of interactions of
quantum objects with a classical device?

The locality problem has become more and more relevant in recent
years. This is because it passes more and more from the domain of
theoretical reasonings and Gedankenexperiments into the domain of
actual experiments. Moreover, the first attempts are being made to
construct prototypes of engineering constructions in the domain of
so-called quantum telecommunication. The quantum locality problem
plays a key role in this domain.

We mention here that the results of modern experiments are
interpreted in most cases as evidence that a "local physical
reality" does not exist in quantum physics. The locality of the
quantum theory then acquires a somewhat metaphysical status
removed from material reality.

Here, we attempt to demonstrate a quantum theory formulation that,
on one hand, preserves the mathematical apparatus of the standard
quantum mechanics and, on the other hand, admits an explicit
interpretation of the locality property.

\section{Basic notions and postulates}

The basic notions of the proposed approach to quantum mechanics
were presented in~\cc{slav1} and described in more detail
in~\cc{slav2}. We only briefly review these notions here. We chose
the inductive method of theory construction in~\cc{slav1}. We
first considered a physical phenomenon, singled out its basic
characteristics, and then described them mathematically.

Here, we choose the deductive method, i.e., we formulate
mathematical postulates from the very beginning. Those who are
interested in their phenomenological justification can turn
to~\cc{slav1} and~\cc{slav2}. The entire construction is performed
in the framework of the algebraic approach. We therefore do not
assume that a physical system state is described by a vector of a
Hilbert space (or by a density matrix) and that observables are
described by operators in this space.

We take the following postulates as basic.

\

\post{} 1. Observables of a physical system are described by
Hermitian elements of some $C^*$-algebra~\AAA.

\

Elements of the algebra \AAA{} are called dynamical variables. We
let~$\AAA_+$ denote the set of observables. We let~$\QQQ$ denote
maximum commutative subalgebras of the algebra~\AAA{} belonging
to~$\AAA_+$. We use the subscript~$\xi\in\Xi$ to distinguish among
these subalgebras.

\

\post{} 2. For observables $\A$ and $\B$ to be compatible
(simultaneously measurable), it is necessary and sufficient for
them both to belong to some subalgebra $\QQQ$.

\

We let $\vx(\cdot)$ denote a character of the subalgebra $\QQQ$,
i.e., $ \A\stackrel{\vx}{\longrightarrow}\vx(\A)$ is a homomorphic
map of the algebra $\QQQ \quad (\A\in \QQQ)$ into the algebra of
real numbers.

We call the collection  $\vp=[\vx] \quad (\xi\in\Xi)$ of
functionals $\vx(\cdot)$, each of which is a character of the
corresponding subalgebra~$\QQQ$, an elementary state of a physical
system.

\

\post{} 3. The result of any individual experiment on measuring
physical system observables is determined by an elementary state
of this system.

\

The same observable may simultaneously belong to several
subalgebras $\QQQ$. We say that an elementary state $\vp$ is
stable on the observable $\A$ if for all subalgebras
$\QQQ,\;\qqq,$ containing $\A$, we have the equality

 \beq{1}
\vx(\A)=\vp_{\xi'}(\A), \mbox{ если }\A \in \QQQ\cap\qqq,\quad
\vx(\cdot)\in\vp,\; \vp_{\xi'}(\cdot)\in\vp.
 \eeq

Condition \rr{1} does not hold in the general case. This means
that the measurement result can depend not only on the elementary
state $\vp$ but also on the index $\xi\in\Xi$. We say that a
device used for measuring the observable $\xi\in\Xi$ is of type
$\xi$ if using this device, we obtain $A_{\xi}=\vx(\A)$ as the
measurement result. We thus set a definite type of measuring
device into correspondence with each subalgebra $\QQQ$. We must
use the  devices to perform the compatible measurements of
observables belonging to the same subalgebra $\QQQ$.

We say that a measurement of an observable $\A$ is reproducible if
a repeat measurement of the same observable (by a device not
necessarily of the same type) produces a result coinciding with
the initial result. Obviously, a reproducible measurement acts on
a physical system by transforming it into an elementary state that
is stable with respect to the observable~$\A$.

Because compatible measurements are possible only for compatible
observables, we cannot unambiguously fix the elementary state
$\vp$ in an experiment. The maximum that we can do is to determine
the reduction of the elementary state to a subalgebra $\QQQ$,
i.e., we can fix the functional~$\vx(\cdot)$. We say that
elementary states $\vp$ are $\vx$-equivalent if they have the same
reduction $\vx(\cdot)$ on the subalgebra ~$\QQQ$.

We define the purely quantum state $\Psi_{\vx}$, to be the class
$\{\vp\}_{\vx}$ of $\vx$-equivalent elementary states that are
stable with respect to the subalgebra~$\QQQ$. We can therefore
experimentally determine only whether a system under investigation
belongs to a definite quantum state.

\

\rem. The above definition of the quantum state is applicable only
to physical systems in which there are no identical particles. If
such particles are present, then we must replace the equivalence
with weak equivalence in the quantum state definition
(see~\cc{slav2}).

\

The collection of physical systems whose elementary states
constitute the equivalence class $\{\vp\}_{\vx}$ is called a pure
quantum ensemble. \

\post{} 4. A quantum ensemble admits the structure of the
probability space.

\

We recall that the probability space is the fundamental object in
the classic probability theory (see, e.g., \cc{kol,nev}). The
probability space is a triple $(\Om,\F, P)$. The first term in the
triple, $\Om$, is a set of elementary events. In our case, the
elementary state $\vp$ plays the role of an elementary event. The
second term of the triple, $\F$, is the Boolean $\sss$-algebra of
the set $\Om$. Elements of the algebra $\F$ are subsets of the set
$\Om$. These subsets are called (probability) events. Among them,
there must be the set $\Om$ itself and the empty set $\emptyset$.
The algebraic operations in $\F$ are the operations of taking the
union of subsets, their intersection, and the complement with
respect to the set $\Om$. The algebra $\F$ must be closed
(invariant) under the operation of taking the complement and a
denumerable number of the union and intersection operations. The
third term in the triple is a probability measure $P$. This is a
map of the set $\F$ into the set of real numbers:   each $F\in\F$
is sent to a number P(F).   This mapping must satisfy the
conditions $0\leq P(F) \leq 1$ for all $F\in\F$, $P(\Om)=1$, and
$P(\sum_j F_j)=\sum_j P(F_j)$ for any denumerable union $\sum_j
F_j$ of nonintersecting subsets $F_j\in \F$.

We recall that the probability measure is defined only for the
events $F\in\F$. For elementary events, the probability measure
may not exist in general. This is because before considering a
possibility of this or that event, we must ensure that this event
can be realized (at least, in principle). Because the resolution
of a measuring device is finite, we cannot single out an
elementary event. It can be demonstrated (see, e.g., \cc{slav2})
that the assumption that a probability measure necessarily exists
for elementary events results in a contradiction in some cases.

In this situation, quantum theory imposes more severe restrictions
than the classical theory. This is because in order to single out
an event, we must have the possibility to perform simultaneous (or
at least compatible) measurements of a collection of observables.
In the quantum case, only observables belonging to the same
subalgebra $\QQQ$ are compatible. Hence, with each such subalgebra
$\QQQ$, we must associate its own $\sss$-algebra $\F$ and its own
system of probability measures $P_{\xi}(F)$, where $F\in\F_{\xi}$.

On the other hand, there are events that can be singled out by
measuring observables belonging to different subalgebras $\QQQ$.
An example of such an event is the event $F_A$ stating that we
register a value not exceeding $A$ in the experiment for an
observable $\A$. If $\A\in\QQQ\cap\qqq$, then this event can be
registered by both type-$\xi$ and type-$\xi'$ devices. From our
experience, we can conclude that the probability of such the event
is independent of the type of the device we use. We must therefore
introduce one more postulate.

\

\post{} 5. If $\A\in\QQQ\cap\qqq$, then for the system in a
quantum state $\Psi$, the probability of the event $F_A$ is
independent of the type of device used, i.e., $P(\vp: \vpx)\le A)=
P(\vp:\vp_{\xi'}(\A)\le A)$.

\

The mathematical representation of a physical system is the
algebra of its dynamical variables; vice versa, the physical
representation of the algebra of dynamical variables is some
physical system. We can therefore consider the physical
representation of a subalgebra to be the corresponding physical
subsystem. This subsystem is by no means isolated from the rest of
the system, i.e., it can be an open system and not have its own
dynamics. But in most cases, the conclusions of the probability
theory are not related to the dynamics. In particular, we can
treat the subalgebra $\QQQ$ as an algebra of observables of a
classical subsystem of the quantum system under investigation.
Because we can confine ourself to the measurements compatible with
the measurements of observables from the subalgebra $\QQQ$ in
order to find the mean $\langle\A\rangle$ of an observable
$\A\in\QQQ$, the classical probability theory suffices for
calculating such a mean. The formula

 \beq{2}
\langle\A\rangle=\int_{\vp\in\Psi}\,P_{\A}(d\vp)\,A(\vp)
\equiv\int_{\vp\in\Psi}\,P_{\A}(d\vp)\,\vp(\A).
  \eeq
then holds. Here,
 \beq{3}
P_{\A}(d\vp)=P(\vp:\vp(\A)\leq A+dA)-P(\vp:\vp(\A)\leq A),\qquad
A_{\xi}(\vp)\equiv\vpx.
  \eeq
Having in mind Postulate 5, we can omit the index $\xi$ of the
functionals $A_{\xi}(\vp)$ and $\vpx$ in formulas \rr{2} and
\rr{3}.

Formula \rr{2} determines the mean of the observable $\A$ with
respect to the quantum ensemble. We can define the quantum mean
$\Psi(\A)$ experimentally as the arithmetic mean of the results of
measurements for the observable $\A$. The relation between the
quantities $\langle\A\rangle$ and $\Psi(\A)$ is established by the
Khinchin theorem (the large number law; see, e.g., \cc{nev}),
which can be formulated as follows in the terms used in this
paper.

\

{\sc Теорема.} $A_{\xi_j}=\vp_{\xi_j}(\A)$ be the result of
measuring an observable $\A$ in the experiment with the number
$j$. Let $A_{\xi_j}$ be random mutually independent quantities
having the same probability distribution $P_{\A}$ with the finite
expectation $\langle\A\rangle$. The quantity
$n^{-1}(A_{\xi_1}+\dots+A_{\xi_n})$ then converges to
$\langle\A\rangle$ in the probability sense as $n\to\infty$.
Therefore,

 \beq{4}
  \Psi(\A)\equiv\lim_{n\to\infty}\mbox{P}\Big[n^{-1}
\Big(\vp_{\xi_1}(\A)+\dots+\vp_{\xi_n}(\A)\Big)\Big] =
\langle\A\rangle.  \eeq

Experiment proves that the following statement holds.

\

\post{} 6. The quantity $\Psi(\A)$ is a linear functional of
observables, i.e.,

 $$
 \Psi(\A)+\Psi(\B)=\Psi(\A+\B)\mbox{ for all }\A,\B\in\AAA_+.
 $$

  \
This functional can be unambiguously extended to the algebra
$\AAA$ using the formula $\Psi(\A+i\B)=\Psi(\A)+i\Psi(\B)$, where
$\A,\B\in\AAA_+$.

Every $C^*$-algebra $\AAA$ is isometrically isomorphic to a
subalgebra $\BB(\HHH)$ of bounded linear functionals in a Hilbert
space $\HHH$ (see, e.g., \cc{dix}), i.e.,

$$
\A\leftrightarrow\Pi(\A), \quad \A\in\AAA, \quad
\Pi(\A)\in\BB(\HHH). $$

It can be shown (see \cc{slav1,slav2}) that the mean
$\langle\A\rangle$ of the observable $\A$ with respect to the
quantum ensemble $\Psi$ defined by formula \rr{2} can be
represented as the expectation of the operator $\Pi(\A)$:

 \beq{6}
 \langle\A\rangle=\langle\Psi|\Pi(\A)|\Psi\rangle,
  \eeq
where $|\Psi\rangle\in\HHH$ is the corresponding vector in the
Hilbert space.

\section{Locality of observables and nonlocality of states}

Formulas \rr{4} and \rr{6} indicate that, on one hand, we can use
the mathematical tools of the standard quantum mechanics to
calculate quantum means $\Psi(\A)$ and, on the other hand, we can
interpret a quantum state as an equivalence class of elementary
states. An equivalence class is a mathematical notion existing out
of time and space. Speaking about a localization of a quantum
state is therefore absurd.

The equivalence class corresponding to a definite quantum state
can be composed using features common to all the elements from
this class. Such common features can be the same values of
observables obtained with a definite measurement procedure.
Neither coordinate nor time are observed quantities. They are
parameters of the Minkowski space in which physical objects dwell.
But an observable called the "coordinate" is very often used in
nonrelativistic quantum mechanics.

To find a way out of this mess, we consider the procedure for
measuring this "coordinate" in more detail. For simplicity, we
assume the physical object under study to be pointlike. To measure
the "coordinate" of a physical object, we use a measuring device,
a ruler with graduations. To each graduation interval, we
associate the observable $\p_i$, where $i$ is the number of the
corresponding interval. We ascribe the value 1 to the observable
$\p_i$ if the physical object under study is inside the $i$-th
interval and the value 0 to the observable $\p_i$ if it is outside
this interval. The observable $\p_i$ thus defined has the
properties of the projector and is an element of the algebra of
dynamical quantities.

Freely speaking, we call the index $i$ the coordinate and state
that the object under study has the coordinate $i$ if the value of
the corresponding observable $\p_i$ is equal to 1. This
"coordinate" $i$ is very distantly related to the genuine
coordinate, which is a parameter of the Minkowski space. We can
repeat our experiment in another domain of the Minkowski space. We
must then carry both the object under study and the ruler to this
domain. If the object under study again turns out to be in the
interval with the number $i$, then we say that the second
experiment produces the same value of the "coordinate."  We can
organize the equivalence class with respect to just this
"coordinate" (in fact, with respect to the value of the observable
$\p_i$)  and say that we have constructed a quantum state
concentrated in the vicinity of the "coordinate" $i$. This
"coordinate" has no relation to a localization in the Minkowski
space.

To discuss the locality problem in more detail, we consider how we
can describe particle scattering by two slits $a$ and $b$ using
the idea of the elementary state. An interference pattern is
vividly observed in this experiment. This picture is clearly
determined by the probability distribution of particle momenta
after scattering. Three events are essential in the experiment
under consideration: the event $F_a$, which means that the
particle hits a domain of slit $a$, the event $F_b$, which means
that the particle hits a domain of slit $b$, and the event $F_k$,
which means that the scattered particle momentum falls into a
fixed small solid angle around the direction $K$.

The problem under consideration can be formulated in these terms
as a typical problem of calculating conditional probability. We
must calculate the probability of the event $P(F_k)$ under the
condition of realization of either event $F_a$ or event $F_b$.
Classical probability theory provides a standard formula, but we
cannot apply it directly in the quantum case because it involves
the probability of simultaneous realization of the events $P(F_k)$
and $F_a+F_b$. But a probability measure does not exist for this
event because the events $P(F_k)$ and $F_a+F_b$ are incompatible
because of the incompatibility of simultaneous measurements of the
coordinate and the momentum.

But we can propose a detour for calculating such a conditional
probability. For this, it suffices to consider the first stage of
scattering in which the particle hits either the domain of slit
$a$ or the domain of slit $b$ as the preparation of a quantum
state. When using this quantum state as the new probability space,
we can consider the event $F_k$ as an unconditional one.

We can set the observable $\p_a$, which takes the value $p_a=1$ if
the particle hits the domain of slit $a$ and value $p_a=0$ if the
particle misses this domain, into correspondence with the event
$F_a$. We set the analogous observable $\p_b$ into correspondence
with the event $F_b$. Only those particles whose elementary states
correspond to the value of the observable $\p_a+\p_b$ equal to one
contribute to the interference pattern. Such elementary states
constitute an equivalence class, denoted by $\Psi_{a+b}$. Because
the observable $\p_a+\p_b$ is not the only independent generator
of the maximum subalgebra of compatible observables in the general
case, the quantum state corresponding to the equivalence class
$\Psi_{a+b}$ can be mixed. But even in this case, the functional
describing the means of observables with respect to this quantum
state is positive definite, linear, and normalized to unity. We
let $\Psi_{a+b}(\cdot)$ denote this functional. It has the
property

 \beq{7}
 \Psi_{a+b}(\I)=1,
 \eeq
where $\I$ is the unit element of the algebra \AAA. Moreover, for
all the elementary states in this quantum ensemble, we have
$p_a+p_b=1$, and the functional $\Psi_{a+b}(\cdot)$ by virtue of
formulas $\Psi_{a+b}(\cdot)$ \rr{2} and \rr{4} satisfies the
condition

 \beq{8}
 \Psi_{a+b}(\p_a+\p_b)=1.
 \eeq

Because the functional $\Psi_{a+b}(\cdot)$ is positive definite,
the Cauchy-Buniakowski-Schwarz inequality holds for it,
 \beq{9}
\lt|\Psi_{a+b}\lt(\A(\I-\p_a-\p_b)\rt)\rt|^2\leq\Psi_{a+b}(\A^*\A)
\Psi_{a+b}(\I-\p_a-\p_b).
 \eeq

By virtue of equalities \rr{7} and \rr{8}, the right-hand side of
inequality \rr{9} is zero. Therefore,

 \beq{10}
\Psi_{a+b}(\A)=\Psi_{a+b}\lt(\A(\p_a+\p_b)\rt).
 \eeq
Analogously,
  \beq{11}
\Psi_{a+b}(\A)=\Psi_{a+b}\lt((\p_a+\p_b)\A\rt).
 \eeq
Replacing $\A\to\A(\p_a+\p_b)$ in \rr{11} and taking \rr{10} into
account,  we obtain
  \beq{12}
\Psi_{a+b}(\A)=\Psi_{a+b}\lt((\p_a+\p_b)\A(\p_a+\p_b)\rt).
 \eeq

We set the observable $\K$ into correspondence with the event
$F_k$. Using formula \rr{12}, we obtain the expression for the
mean of this observable

  \beq{13}
\langle\K\rangle=\Psi_{a+b}(\K)=
\Psi_{a+b}(\p_a\K\p_a)+\Psi_{a+b}(\p_b\K\p_b)+
\Psi_{a+b}(\p_a\K\p_b+\p_b\K\p_a).
 \eeq
The first and second terms in the right-hand side of \rr{13}
describe the scattering from the respective slits $a$ and $b$. The
third term describes the interference. Because
$\p_a\p_b=\p_b\p_a=0$, in the case where $[\p_a,\K]=0$ or
$[\p_b,\K]=0$, the interference term disappears.

The interference pattern is purely determined by the structure of
the abstract equivalence class $\Psi_{a+b}$ would therefore
organize our experiment as follows. We can prepare many copies of
the same experimental device and distribute it over the globe. At
each device, we perform one scattering act at random time
instants. We then put together all the screens on which we have
spots from hits of the scattered particles and put all these
screens in one stack. For a sufficiently large number of screens,
we must obtain a pattern close to that described by
formula~\rr{13}.

We note that in contrast to considering the same experiment in the
standard quantum mechanics, we consider that the scattered
particle hit either the domain of slit $a$ or the domain of slit
$b$ in each separate case, not passing in a mysterious way through
both slits simultaneously. This means that we consider a particle
well localized in each separate act. The interference pattern
appears because the functional $\Psi_{a+b}(\cdot)$ cannot be
represented as a sum of the functionals $\Psi_{a}(\cdot)$ and
$\Psi_{b}(\cdot)$ corresponding to the respective acts of separate
scattering on the slits $a$ and $b$. Physically, this means that
the scattering on one slit depends on the presence or absence of
the other slit, i.e., a nonlocality is present here.

We see how this nonlocality can be explained in the framework of a
local field theory. For clarity, we here discuss the example of
the process of scattering of an electron on a nucleus, well
studied both theoretically and experimentally. Because the
electron is much lighter than the nucleus, this process is well
approximated by the electron scattering on a classical source. In
what follows, we discuss exactly this process in the framework of
the perturbation theory in the standard quantum electrodynamics
(see, e.g., \cc{pesk}).

In the first order of the perturbation theory in the electron
charge, this process is described by the Feynman diagram (a) shown
in Fig. 1. In this figure, straight lines correspond to the
electron, wavy lines correspond to the photon, and the crossed
circle corresponds to the source of the classical electromagnetic
field. Calculating the differential scattering cross section when
taking diagram (a) into account causes no troubles and results in
the celebrated Rutherford formula corrected by taking the electron
spin into account. The obtained formula describes the experimental
situation well. But both theory and experiment have now gone far
beyond the accuracy level ensured by the first correction to the
perturbation theory.

 \begin{figure}[h]
 \begin{center}

  \includegraphics{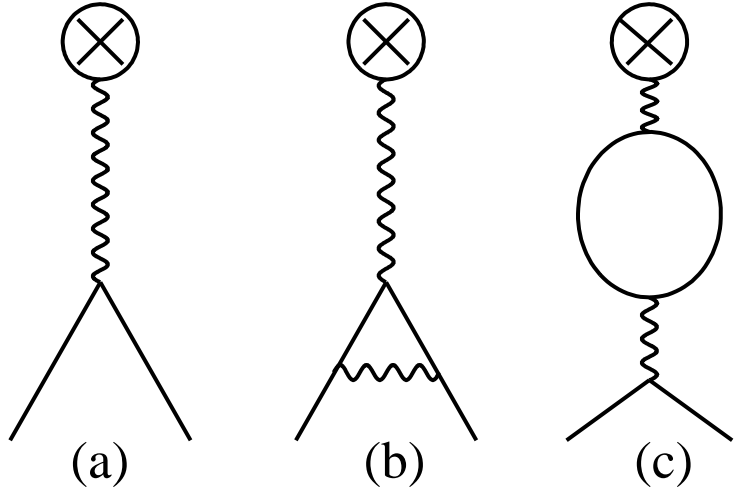}

   \caption{Diagrams describing the elastic scattering.}
\end{center}
\end{figure}

The next order of the perturbation expansion that contributes to
the process under study is the third order. There, we must take
contributions coming from diagrams (b) and (с) in Fig. 1 into
account. Taking these diagrams into account results in substantial
theoretical difficulties. First, the so-called ultraviolet
divergences appear because the intermediate (virtual) particles
can carry arbitrarily large energies and momenta. In quantum field
theory, a well-defined algorithm (the renormalization theory) was
developed to overcome this difficulty.  We do not discuss this
problem in what follows.  Second, diagram (b) results in the
so-called infrared divergences caused by the presence of massless
particles in the complete particle set. In the example under
consideration, such particles are the photons. Quantum field
theory also provides an algorithm for overcoming this difficulty.
We discuss it in more detail.

The algorithm is based on the following experimental fact. The
elastic scattering process described by diagrams (a-c) cannot be
experimentally separated from the process of bremsstrahlung
depicted in diagrams (d-f) in Fig. 2. In this process, electron
scattering is accompanied by emitting one (diagrams (d) and (e))
or more (diagram (f)) photons. The contribution of diagrams of
such type to the scattering cross section cannot be experimentally
separated from the contributions of diagrams (a-c) if the total
energy of photons emitted in the bremsstrahlung is below the
sensitivity threshold of the measuring device. \\ [-2cm]

\begin{figure}[h]
 \begin{center}

  \includegraphics{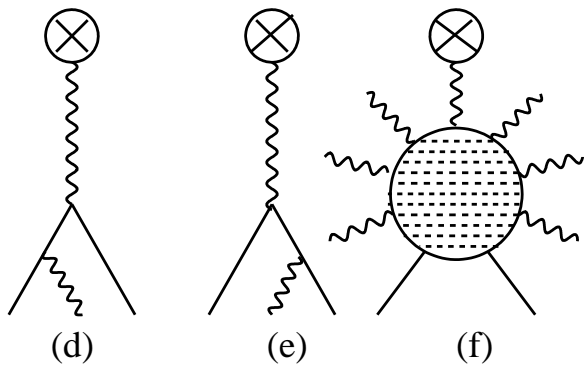}

   \caption{Diagrams describing the bremsstrahlung.}
\end{center}
\end{figure}

Calculations show that if we take diagrams (d) and (e) into
account together with diagrams (a-c), then infrared singularities
are compensated. But the scattering cross section then becomes
dependent on a parameter characterizing the sensitivity of the
measuring device. We can take the total energy $E_{max}$ of
additionally emitted photons as such a parameter. This is an
absolutely physical parameter, and the dependence of the measured
scattering cross section on this parameter should therefore not
cause any principal objection. But taking diagrams (a-e) alone
into account results in one more difficulty. The dependence of the
scattering cross section on the parameter $E_{max}$ is singular,
and the cross section can become negative at sufficiently small
$E_{max}$.

In modern theory, this difficulty is attributed to an artifact
related to using the perturbation theory. Indeed, if higher orders
of the perturbation theory are considered taking contributions
from diagrams of type (f) with an infinitely increasing number of
emitted photons into account and summing all these contributions,
then the cross section dependence on the parameter $E_{max}$
becomes regular.  Moreover, this cross section tends to zero as
$E_{max}\to 0$. This does not cause objections from the physical
standpoint. The purely elastic scattering in which no
bremsstrahlung photons are emitted is just one among infinitely
many channels along which this process may proceed. It is
therefore not amazing that each of these channels contributes
infinitesimally to the total cross section.

Here we need a new insight into the process called elastic
scattering. In reality, this process is never purely elastic.
Electron scattering is always accompanied by the bremsstrahlung,
which cannot be registered even by a measuring device with very
high sensitivity. Moreover, the result of the experiment becomes
strongly dependent on the device sensitivity if the latter is too
high. When the sensitivity becomes infinitely high, the registered
scattering cross section must tend to zero because we study the
measuring device in this case and not the physical object
(electron) under investigation.

The above example teaches us several useful lessons.

Lesson 1. Separating characteristics of a physical object under
study is somewhat conditional. These characteristics cannot be
separated completely from the characteristics of the measuring
device with which this object interacts.

Lesson 2. A physical object (electron in the above example) with
which we associate definite physical characteristics (differential
cross section of elastic scattering) is accompanied by a field
(bremsstrahlung photons) that is not registered by the measuring
device but affects the result of the measurement of the
characteristics under study.

Lesson 3. The presence of the accompanying field does not
contradict locality axioms of quantum field theory. In the above
example, both the electron and the bremsstrahlung photons
propagate in the future light cone with the vertex at the
scattering point.

Lesson 4. The result of measuring the characteristics (scattering
cross section) ascribed to a well-localized object under study
(electron) may depend on the characteristics of physical objects
(bremsstrahlung photons) that are located in the domain that is
spacelike with respect to the localization domain of the object
under study. This may be interpreted as a nonlocality of the
object under study.

The above lessons result in the following conclusion. We can split
the physical problem under study into two parts as regards the
measurement process. The first part, called the {\it kernel} in
what follows, is registered by a measuring device. The second
part, called the {\it dark field} in what follows, is not directly
registered by the measuring device, but the instrument reading can
depend on characteristics of the dark field. The separation into
these two parts is not absolute and depends on the measurement
procedure. This mobility of the boundary finds its partial
realization in the renormalization group formalism in the
mathematical apparatus of quantum field theory.

In the framework of the algebraic approach to quantum theory
described in this paper, such a division of a physical system into
two parts can be related to two elements of the mathematical
apparatus: the algebra of dynamical quantities and the elementary
state. Each quantum particle reveals itself through the
corresponding observable quantities or, more precisely, through
local observables whose values can be found by performing
measurements in a bounded domain $\OO$ of the Minkowski space. We
can therefore consider the local algebra $\AAA(\OO)$ to be the
mathematical representation of the quantum particle. In the
standard algebraic approach to quantum field theory, this algebra
is customarily called the algebra of local observables. This is
not completely correct because the set of observables constitutes
not an algebra but rather a subset of the local algebra of the
corresponding dynamical quantities.

The domain $\OO$ can be naturally considered to be a localization
domain for the quantum particle under consideration.  In any case,
the domain $\OO$ must contain the particle localization domain.
More precisely, the localization domain of a particle must be
associated with its kernel. Indeed, as explained above, the
registered values of observables may depend on the characteristics
of the dark field. This field is not necessarily localized in the
domain $\OO$. On the other hand, in the approach of this paper,
observable values are determined by the elementary state. The
matter carrier of the elementary state is therefore not only the
kernel of the quantum object under study but also the associated
dark field. The elementary state therefore cannot be regarded as
being localized in the domain $\OO$. But in contrast to the
quantum state, which has no localization in the Minkowski space,
the elementary state has a localization: it is the dark field
localization.

We note that not every dark field affects the result of measuring
observables (the scattering cross section in the above example);
only the one that is created together with the kernel does, i.e.,
only the dark field coherent to the kernel is essential.

We now return to discussing the experiment on the scattering from
two slits. We regard the electron as the scattered particle. We
obtained formula \rr{13}, which describes the interference
phenomenon using the fact that the quantum ensemble $\Psi_{a+b}$
has a definite structure. First, at the instant of the ensemble
creation, the electron is localized either in the domain of slit
$a$ or in the domain of slit $b$ (it is better to speak about the
electron kernel, not the electron itself). Second, as a result of
interaction between the electron and the slits $a$ and $b$, an
ensemble is created whose structure is such that the corresponding
functional $\Psi_{a+b}(\cdot)$ is linear.

We in fact replaced a real description of interaction between the
electron and slits by the appropriate boundary conditions. These
boundary conditions suffice for the mathematical description of
the phenomenon under investigation. But it would be desirable to
find out, at least on a qualitative level, what the physical
processes underlying these boundary conditions are.

The first condition for electron localization is self-evident and
does not need additional comments. We only note that it will
certainly provoke frantic objections from orthodox followers of
the standard quantum mechanics, who will insist on that we cannot
speak about electron localization before performing a measurement.
Why not? Only because they cannot say anything meaningful on this
subject?

To explain the second condition qualitatively, we can propose the
following model of how the electron interacts with the slits or,
more precisely, with the screen in which these silts are cut. In
the scattering process, not only the electron kernel but also the
companion dark field that is coherent with the kernel approach the
screen. Because this field is massless, it reaches the screen even
before the kernel. This field generates collective oscillations of
the screen that are also coherent with the kernel. The arising
oscillations are very weak, but because of the coherence, they may
interact resonantly with the kernel. At least, they may play the
role of a random force participating in creating the probability
distribution of the scattered electron momentum. In contrast to
the electron kernel, the dark field reaches both slits, and the
character of the random force depends essentially on whether only
one slit is open or both slits are open simultaneously. This can
be a physical reason for the appearance of the interference
pattern.

The dark field reveals itself in the experiment when measuring
values of the observables describing the kernel coherent with this
field. Separated from its kernel, the dark field becomes
explicitly experimentally unobservable. It is therefore a good
candidate for the role of a constituent of dark matter. Of course,
in addition to the electromagnetic field, other massless fields
such as gluon and gravitational fields may contribute to the dark
field.

As stated above, to remove infrared divergences self-consistently,
we must assume, for example, that in electromagnetic interactions,
an infinite number of photons with a finite total energy is
emitted. Such a system behaves as a classical electromagnetic
field, and massless observable quantum fields must therefore
feature classical "tails." This opens an interesting perspective
for a gravitational field. A consistent quantum model of a
gravitational field is still missing despite numerous attempts to
construct it.  The quanta of a gravitational field, the gravitons,
have never been observed. Can it be that the gravitational field
consists only of the classical "tail?"

The dark field mechanism provides a very clear explanation of the
result of the experiment with the so-called delayed choice.
Wheeler proposed the idea of this experiment 30 years ago
\cc{wheel}. Wheeler's idea was recently realized almost
ideally~\cc{jac}. The actual experiment completely confirmed
Wheeler's predictions.

\begin{center}
\begin{picture}(110,80)
\thicklines
 \put(5,30){\vector(1,0){15}}
\put(5,60){\vector(1,0){15}} \put(20,30){\vector(1,0){25}}
\put(5,65){\vector(0,-1){20}} \put(5,45){\line(0,-1){15}}
\put(35,60){\vector(0,-1){15}} \put(5,70){\vector(0,-1){5}}
\put(35,45){\vector(0,-1){25}}
 \put(20,60){\line(1,0){15}}
\put(35,20){\oval(5,7)} \put(45,30){\oval(7,5)}
\put(5,60){\circle{6}} \put(7,55){$M_1$} \put(27,55){$M_2$}
\put(7,33){$M_3$} \put(25,20){$D_B$} \put(43,35){$D_A$}
 \put(17,62){$B\rightarrow$}  \put(7,45){$A\downarrow$}
 \put(17,10){$(a)$}

 \put(60,30){\vector(1,0){15}} \put(60,60){\vector(1,0){15}}
\put(75,30){\vector(1,0){15}}
 \put(60,65){\vector(0,-1){20}}
\put(60,45){\line(0,-1){15}} \put(90,60){\vector(0,-1){15}}
\put(60,70){\vector(0,-1){5}} \put(90,45){\vector(0,-1){25}}
 \put(75,60){\line(1,0){15}}
\put(90,20){\oval(5,7)} \put(100,30){\oval(7,5)}
\put(60,60){\circle{6}} \put(90,30){\circle{6}} \put(62,55){$M_1$}
\put(82,55){$M_2$} \put(62,33){$M_3$} \put(82,33){$M_4$}
\put(80,20){$D_B$} \put(98,35){$D_A$}
 \put(72,62){$B\rightarrow$}  \put(62,45){$A\downarrow$}
 \put(72,10){(b)}
 \thicklines
\put(3.1,32){\line(1,-1){4}} \put(3.1,61.9){\line(1,-1){4}}
 \put(33,61.9){\line(1,-1){4}}

\put(57.9,32){\line(1,-1){4}} \put(57.9,61.9){\line(1,-1){4}}
\put(88,32){\line(1,-1){4}} \put(88,61.9){\line(1,-1){4}}
\end{picture}

Figure 3.      The delayed choice experiment.
\end{center}

The principal scheme of the experimental setup is depicted in Fig.
3. In this figure, $M_2$ and $M_3$ are two totally reflecting
mirrors, and $M_1$ and $M_4$ are two half-silvered mirrors. Mirror
$M_4$ is removable. By the experimenter's choice, it can be either
absent (the device is then in position $(a)$) or present (the
device is then in position $(b)$)). Single photons are emitted
toward mirror $M_1$ at time intervals such that no more than one
photon can be in the device at each given instant. After passing
through the device, the photon reaches either detector $D_A$ or
detector $D_B$

If the device is in position $(a)$ and photon behaves as a
particle, then after passing through mirror $M_1$ it chooses the
path $A$ or $B$ with equal probabilities. As a result, it reaches
either detector $D_A$ or detector $D_B$.

If the device is in position $(b)$ and photon behaves as a wave,
then the process of passing through the installation can be
described as follows. The photon-wave reaches mirror $M_1$. The
wave here splits into two coherent parts. One part propagates
along path $A$, and the other part propagates along path $B$. The
coherence of the parts is preserved. The wave phase is changed by
$\pi/2$ when the wave is reflected by any of the mirrors, and the
phase remains unchanged when the wave passes through a mirror. The
coherent addition of the two waves occurs at mirror $M_4$. The
balance of the phase changes is such that the wave propagates only
in the direction of detector $D_B$ after reaching mirror $M_4$.

The result of the actual experiment is as follows. If the device
is in position (a), then detector $D_A$ responds with probability
0.5, and detector $D_B$ responds with the same probability. If the
device is in position b, then detector $D_B$ responds with
probability 1. Therefore, in accordance with the device position,
the photon behaves either as a particle or as a wave. Such photon
behavior is agrees with Bohr's context principle~\cc{bohr2}.
According to this principle, the result of a quantum experiment
depends on the general context of the experiment.

But Wheeler proposed complicating the choice problem for the
photon. He proposed installing or removing mirror $M_4$ after the
photon has passed through the mirror $M_1$, i.e., the photon must
predict the subsequent acts of the experimenter. The actual
experiment demonstrated that a photon handles this task
successfully and behaves properly in every situation: either as a
particle or as a wave.

It seems that a time nonlocality is manifested in this experiment:
the future action (the manipulation of mirror $M_4$) affects the
preceding action (the photon's choice to behave as a wave or as a
particle). Wheeler himself interpreted the result of this
experiment confirming the principle that "no registration of the
experimental result means that no physical phenomenon exists."

Explaining the experimental result is much simpler in the
framework of the dark field mechanism. When the photon interacts
with mirror $M_1$, in addition to the scattering (reflection or
passage) of the photon, bremsstrahlung photons are created. The
scattered photon kernel propagates along either path $A$ or path
$B$. The bremsstrahlung photons (the dark field) propagate along
both paths. Both parts of the dark field reach mirror $M_4$ (if it
is present) where they are added coherently and generate small
collective oscillations in mirror $M_4$. These small oscillations
are coherent with the photon kernel, interact with it resonantly,
and play the role of a random force directing the kernel toward
detector $D_B$. If mirror $M_4$ is absent, then the photon kernel
propagates along one of the paths it takes in mirror $M_1$ and
reaches either detector $D_A$ or detector $D_B$. No time
nonlocality arises in this case.

The quantum correlation problem is closely related to the
nonlocality problem. This is because these correlations often look
like a distant action. A typical example is the
Einstein-Podolsky-Rosen paradox~\cc{epr}. For example, in the
variant proposed by Bohm~\cc{bom}, the result of measuring the
projection of a spin of one particle from the singlet pair of
particles with the spins 1/2 on one direction instantly and
unambiguously predicts the result of measuring the projection of
the other particle spin on the same direction even if the
particles are separated by a large distance in space. It seems
that this result contradicts the locality principle. But this
contradiction arises only if we assume that the correlation
results from the interaction between the particles at the instant
of the measurement.

The notion that a correlation between separate elements of a
physical system is always due to interaction between these
elements is a deeply rooted delusion. It is even reflected in the
terminology used in quantum mechanics. We can often hear the terms
"exchange interaction," "nonforce quantum action," or reasonings
about "strong quantum correlations."

In fact, quantum correlations are not caused by features specific
to quantum interactions. For example, in the
Einstein-Podolsky-Rosen paradox, the correlations between spin
projections of two particles arise because these particles were
created as a singlet pair for which the law of conservation of the
proper angular momentum is satisfied. But the angular momentum
conservation law also holds in classical physics.

In most cases, quantum correlations are due to the structure of
the physical system ensemble participating in the quantum
experiments. This structure is fixed by the procedure for
preparing the ensemble under consideration, and the preparation
procedure is in turn determined by the properties of the classical
device used. As a rule, quantum correlations are therefore caused
by the interaction between each separate constituent of the
quantum ensemble and the classical device (or devices) preparing
this quantum ensemble, not by the interaction between quantum
objects. This interaction can be smeared both in time and space
for separate constituents of the ensemble. It is therefore not
amazing that it often seems that correlations contradict the
principle of the locality of interaction. In fact, the locality
principle is always satisfied for correlations. But this
correlation must be verified not from the standpoint of
interaction between different constituents of the quantum ensemble
but from the standpoint of interaction of separate constituents of
this ensemble with the devices preparing this ensemble.

\section{Conclusion}

Summarizing, we can draw the following conclusions.

Quantum theory, both relativistic and nonrelativistic, can be
formulated such that it does not contradict the locality condition
accepted in quantum field theory. The measurement process also
does not contradict this condition.

The incompleteness of quantum mechanics noted by Einstein can be
removed by introducing a new notion of the "elementary state,"
which is to be attributed to an individual physical system

From the measurement standpoint, a physical system under study can
be separated into two parts: the so-called kernel and the
accompanying dark field. The kernel is the material carrier of
corpuscular properties of the physical system. The kernel is
localized in the Minkowski space. The algebra of local observables
is the mathematical representation of the kernel. The structure of
the dark field does not contradict the relativistic condition of
locality, but the dark field has a localization worse than that of
the kernel. The elementary state of a physical system is
determined by both the kernel and the dark field structure. The
elementary state is the mathematical representation of the
material carrier of the wave properties of the physical system.

A quantum state is an equivalence class in the set of elementary
states and plays the role of the mathematical representation of
the (quantum) ensemble of physical systems under investigation.
The quantum state does not have the property of locality in the
Minkowski space.

\end{document}